# Conical spin order with chiral quadrupole helix in CsCuCl$_3$

Hiroki Ueda[1,†,*], Elizabeth Skoropata[1], Max Burian[1], Victor Ukleev[1], Gérard Sylvester Perren[1], Ludmila Leroy[1], Julien Zaccaro[2], and Urs Staub[1,*]

[1] *Swiss Light Source, Paul Scherrer Institute, 5232 Villigen-PSI, Switzerland.*

[2] *Grenoble Alpes University, CNRS, Grenoble INP, Institut Néel, BP166, 38042 Grenoble Cedex 09, France.*

**Abstract**: Here we report a resonant x-ray diffraction (RXD) study at the Cu $L_3$ edge on the multi-chiral system CsCuCl$_3$, exhibiting helical magnetic order in a chiral crystal structure. RXD is a powerful technique to disentangle electronic degrees of freedom due to its sensitivity to electric monopoles (charge), magnetic dipoles (spin), and electric quadrupoles (orbital). We characterize electric quadrupole moments around Cu ascribed to the unoccupied Cu 3$d$ orbital, whose quantization axis is off the basal plane. Detailed investigation of magnetic reflections reveals additional sinusoidal modulations along the principal axis superimposed on the reported helical structure, i.e., a longitudinal conical (helical-butterfly) structure. The out-of-plane modulations imply significant spin-orbit interaction despite $S = 1/2$ of Cu$^{2+}$.

† Present address: SwissFEL, Paul Scherrer Institute, 5232 Villigen-PSI, Switzerland.

* To whom correspondence should be addressed: hiroki.ueda@psi.ch and urs.staub@psi.ch

Magnetism and associated functionalities in non-centrosymmetric materials have attracted significant interest in the field of condensed matter physics. These interests lie, for example, in symmetry-protected spin textures, such as skyrmion lattices [Mühlbauer1] and chiral soliton lattices [Togawa1], and in non-reciprocal responses of quantum (quasi-)particles [Tokura1]. The low crystal symmetry is essential to stabilize a complex magnetic ground state with enriched properties due to additional interactions absent in centrosymmetric materials [Dzyaloshinsky1,Moriya1]. On the other hand, the low symmetry adds complexity in solving the magnetic ground state.

Resonant x-ray diffraction (RXD) has been used to explore complex electronic ordered states, e.g., charge, magnetic, or orbital modulations, of which some show chiral orders [Gibbs1,Murakami1,Lang1,Tanaka1]. RXD is based on the anisotropic scattering of x-rays at an atomic resonance, with contributions that are described by tensors up to the second-rank multipole moments $\langle T_Q^K \rangle$ ($-K \leq Q \leq K$), electric monopole ($K = 0$), magnetic dipole ($K = 1$), and electric quadrupole ($K = 2$) [Matteo1]. Here we restrict our interpretation to the electric dipole-electric dipole channel of scatterings, generally most relevant in RXD. An electric monopole corresponds to a charge (spherical electron density), a magnetic dipole corresponds to a magnetic moment, and an electric quadrupole corresponds to an aspheric electron density due to partial electron occupancy of orbital(s) and/or covalency. Therefore,

RXD is a powerful technique for investigating an electronic ordered state of charges, spins, and/or orbitals. Furthermore, the magnetic scattering cross-section can be significant even for materials with small magnetic moments.

Through direct measurements of orbitals and magnetic moments by RXD, we investigated the correlation between the two electronic degrees of freedom in a hexagonal chiral crystal $CsCuCl_3$ with $S = 1/2$. We observed an out-of-plane component of spins in addition to the reported in-plane spin-spiral structure by neutron diffraction [Adachi1], indicating magnetic anisotropy via spin-orbit interaction. Although $S = 1/2$ systems have basically negligible magnetic anisotropy via spin-orbit coupling to their ground state [Moriya2], a recent theoretical study revealed the importance of single-ion anisotropy in some $Cu^{2+}$ based compounds [Liu1]. The here obtained magnetic structure is consistent with those allowed by a symmetry analysis based on group theory. Our results show the powerful potential of RXD for non-centrosymmetric materials with a complex electronic order and a strong correlation between magnetic moments and orbitals even in $S = 1/2$ systems.

$CsCuCl_3$ possesses a distorted hexagonal perovskite structure because of the cooperative Jahn-Teller effect. The room-temperature structure belongs to a chiral space group, either $P6_522$ [left-handed, Fig. 1(a)] or $P6_122$ [right-handed, Fig. 1(b)], that appears below a phase transition temperature of ~423 K [Hirotsu1]. $Cu^{2+}$ with $S = 1/2$ and the Wyckoff position $6a$ form a chiral chain along the principal axis and a triangular lattice in the basal plane, stabilizing a 120° antiferromagnetic (AFM) structure below $T_N$ (= 10.7 K). Intra-chain ferromagnetic exchange interaction and antisymmetric exchange (Dzyaloshinskii-Moriya) interaction, allowed by the low-symmetry, twist the 120° AFM structure along [001] with a periodicity of ~21 nm. The magnetic propagation vector **k** of the helical structure is (1/3, 1/3, $\delta$), where $\delta \approx 0.085$. This magnetic structure reported by neutron diffraction [Adachi1] resembles those of chiral langasite $Ba_3(Nb,Ta)Fe_3Si_2O_{14}$ [Marty1] and double molybdate $RbFe(MoO_4)_2$ [Kenzelmann1]. The former possesses additional sinusoidal modulations of spins along [001], a so-called longitudinal conical (or helical-butterfly) structure [Scagnoli1].

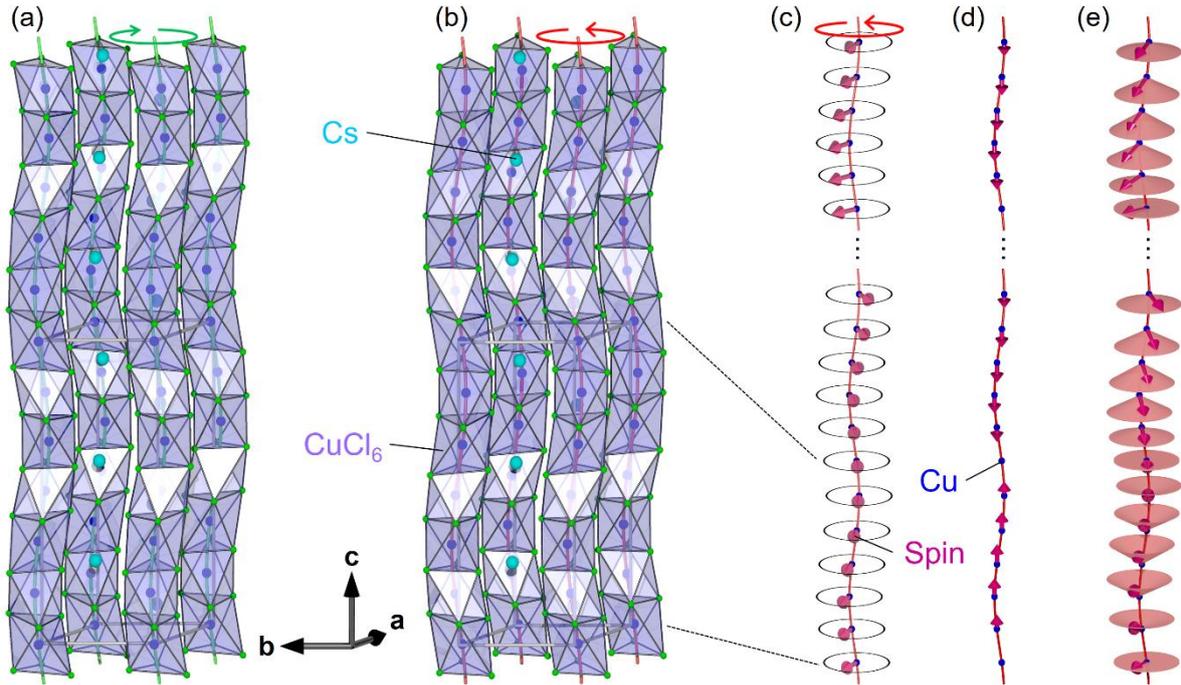

**Fig. 1** Crystal structures of CsCuCl$_3$ [(a) left-handed ($P6_522$) and (b) right-handed ($P6_122$)] and (e) its longitudinal conical (or helical-butterfly) magnetic structure with two components, (c) a helical one parallel to the basal plane and (d) a sinusoidal one parallel to the principal axis. Red and green helices are guides to the eyes for the chiral arrangements of Cu$^{2+}$ along the [001] axis.

As a result of the two-fold ($C_2$) symmetry breaking along <110> reflected by the small $z$ component of **k**, two propagation vectors of **k**$_1$ = (1/3, 1/3, $\delta$) and **k**$_2$ = (1/3, 1/3, –$\delta$) do not coexist in a single magnetic domain, shown in Fig. 2. Such domains characterized by the star of **k** are called configuration domains [Brown1]. Since the helical component gives chirality domains, there are four possible magnetic domains in the reported magnetic structure of CsCuCl$_3$, as shown in Fig. 2.

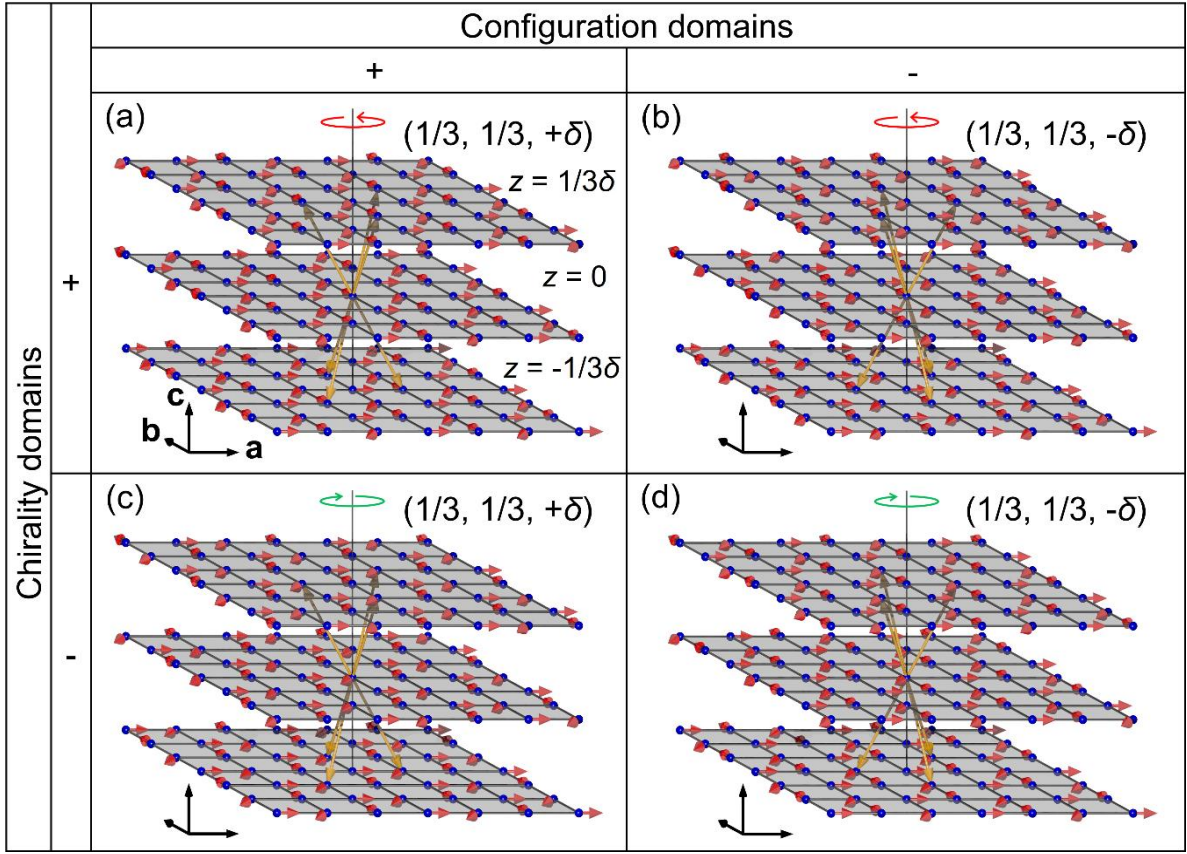

**Fig. 2** The reported magnetic structure of CsCuCl$_3$, and possible four domains. A blue sphere, red arrow, and orange arrow represent Cu$^{2+}$, spin moment, and magnetic propagation vector **k**, respectively. There are six equivalent **k** for respective domains, the star of **k**$_1$ (1/3, 1/3, +$\delta$) or **k**$_2$ (1/3, 1/3, –$\delta$). A gray plane represents the pseudo lattice plane normal to [001] at a different height ($z$).

RXD on CsCuCl$_3$ was previously performed at the Cu $K$ edge ($1s \rightarrow 4p$), and the chiral crystal structure was characterized through the observation of electric quadrupole moments of Cu $4p$ [Kousaka1]. As a result of RXD combined with polarized neutron diffraction, a strong correlation between crystal chirality and magnetic chirality was reported [Kousaka2]. However, RXD at the Cu $L_3$ edge ($2p \rightarrow 3d$) is directly sensitive to $3d$ states, the fundamental orbitals that closely relate to the electronic degrees of freedom of the material, i.e., both magnetism and unoccupied Cu $3d$ orbital ($3d_{x^2-y^2}$), namely, a hole [Laiho1]. Here the quantization axis $z$ is along the elongated direction of CuCl$_6$ octahedron [see Fig. 3(d)]. Such an experiment enables us to investigate the two different orders simultaneously and directly with similar penetration depths, providing ideal comparison conditions.

Our RXD experiments were performed on mono-chiral single crystals probing different surfaces, sample#1: parallel to (001), sample#2: parallel to (119), and sample#3: parallel to (110). These samples were grown from aqueous solution by a method slightly different from Ref. [Kousaka3]. To insure mono-chirality, crystallization was finely controlled, not by evaporation but by a slow temperature lowering of the solution from 40 °C to about 25 °C. We mounted the samples on a diffractometer installed at the RESOXS end-station [Staub1]. The photon energy was chosen around the Cu $L_3$ edge (~930 eV), and the polarization of x-ray beams, linear π and circular C+/C–, was set by the twin Apple II type undulators of the X11MA beamline at the Swiss Light Source (Switzerland) [Flechsig1]. Here C+ (C–) is defined by the Stokes parameter $P_2$ = +1 (–1) [Landau1].

Let us first formulate RXD structure factors to obtain the intensities of the (001) and (002) forbidden reflections, $I_{(001)}$ and $I_{(002)}$, using electric quadrupole moments, $Q_{\xi\eta}$, $Q_{\eta\zeta}$, $Q_{\zeta\xi}$, $Q_{\xi^2-\eta^2}$, and $Q_{3\zeta^2-r^2}$ (see Appendix A for detailed calculation). Here we use a local Cartesian coordinate system $\xi\eta\zeta$ shown in Fig. 3(d). Because of the $C_2$ symmetry along $\xi$, $Q_{\xi\eta}$ and $Q_{\zeta\xi}$ are constrained to be zero. $Q_{3\zeta^2-r^2}$ contributes to allowed Bragg reflections, however, none of which are accessible at the Cu $L_3$ edge. Then $I_{(001)}$ and $I_{(002)}$ are obtained as

$$I_{(001)}^{\pi} = I_{(001)}^{C+} = I_{(001)}^{C-} = \frac{27}{4}|Q_{\eta\zeta}|^2\cos^2\theta, \qquad (1)$$

$$I_{(002)}(\chi, P_2) = \frac{27}{8}|Q_{\xi^2-\eta^2}|^2(1+\sin^2\theta)(1-\chi P_2 \sin\theta)^2, \qquad (2)$$

where $\theta$ is the Bragg angle [~21.7° for (001) and ~47.8° for (002)] and $\chi$ is the crystal chirality [–1 (+1) for left- (right-)handed structure]. Hence, we expect that the (002) reflection exhibits circular dichroism corresponding to the handedness of the crystal structure, whereas the (001) reflection does not. These two reflections probe different quadrupole moments shown in Fig. 3(d), in contrast to trigonal chiral crystals [Lovesey1,Usui1], where two quadrupole moments contribute to a forbidden reflection.

Figures 3(a)-3(c) show the RXD profiles taken around the two reflections at the Cu $L_3$ edge, nicely matching with Eqs. (1) and (2). Their resonant enhancement is confirmed by photon-energy scans while fixing the reflection condition, as seen in Figs. 4(a) and 4(b). Here a dip structure significant for (001) around 930.3 eV is due to self-absorption [Scagnoli2,Joly1]. A minor effect is observed for the (002) reflection implying that the "self-absorption" is more significant close to the surface as the (001) is more surface sensitive due to the shallower incident and exit angles than for the (002) reflection. There is enormous circular dichroism on (002) with an intensity ratio of ~50, close to the expected ratio of ~45 obtained by evaluating Eq. (2). The dichroism is indeed negligible on (001), as expected. The relative magnitude of $Q_{\xi^2-\eta^2}$ and $Q_{\eta\zeta}$ is derived by fitting the experimental results using Eqs. (1) and (2). The self-absorption effect was thereby accounted for and determined by fitting the photon-energy scan. The obtained ratio is $|Q_{\xi^2-\eta^2}|:|Q_{\eta\zeta}| \approx 0.8:1.0$. A linear combination of $Q_{\xi^2-\eta^2}$ and $Q_{\eta\zeta}$ with the obtained ratio provides the exact aspheric electron density due to the presence of a hole in $3d_{x^2-y^2}$ as reported in Ref. [Laiho1]. Since left-handed and right-handed structures are connected by a mirror operation in the $\xi\zeta$ plane, $Q_{\eta\zeta}$ flips its sign while $Q_{\xi^2-\eta^2}$ does not, as sketched in Fig. 3(d). As a result, the local electric quadrupole moments form a chiral helix along [001] and are mirrored between the two structures, as observed in a trigonal chiral crystal $DyFe_3(BO_3)_4$ [Usui1].

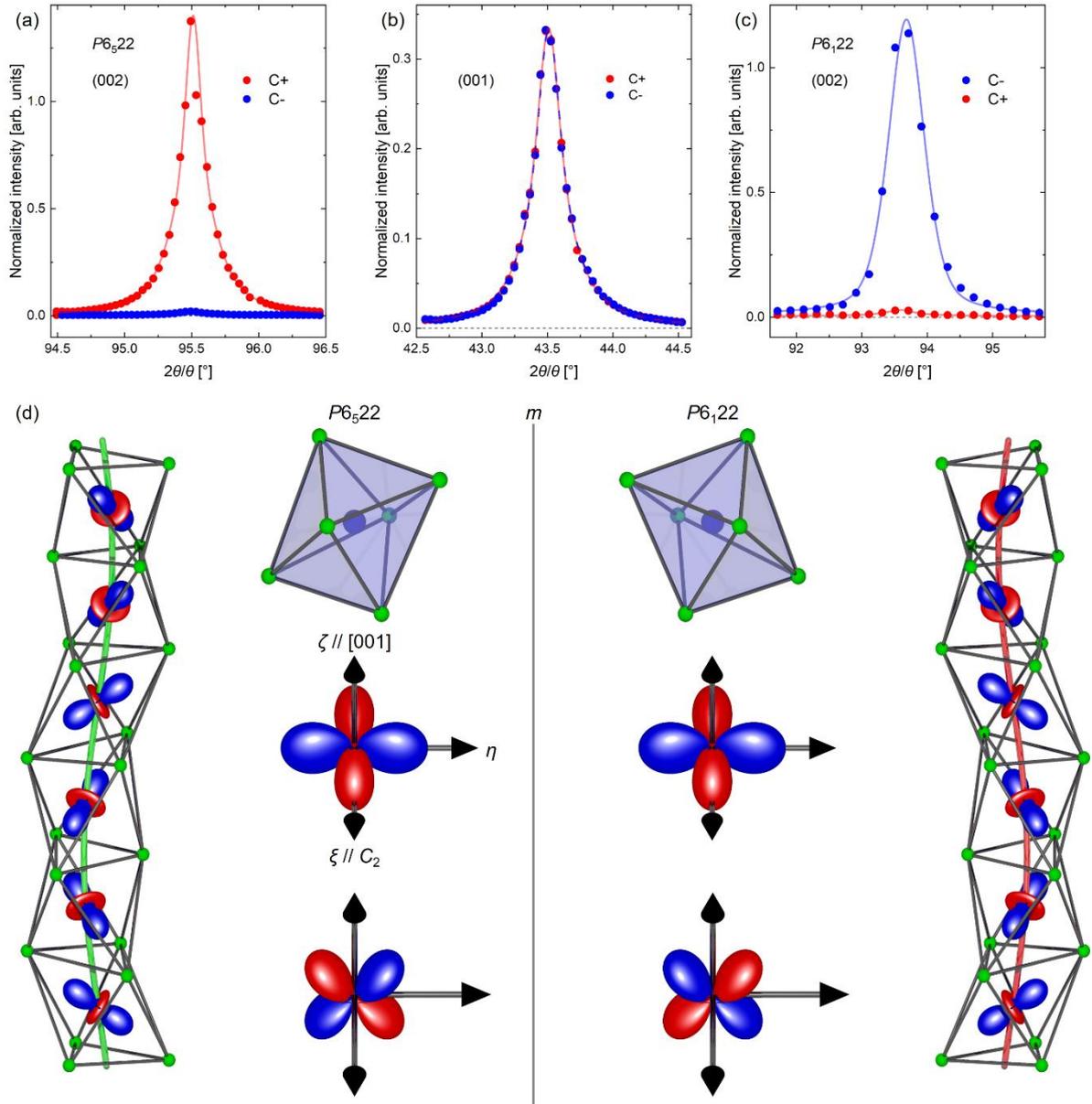

**Fig. 3** Resonant diffraction profiles; (a),(c) around the (002) reflection, where (a) was measured for the left-handed crystal ($P6_522$) while (c) was measured for the right-handed crystal ($P6_122$), and (b) around the (001) reflection. Note that (a) and (b) were taken from sample#1 with the (001) surface, whereas (c) was taken from sample#2 with the (119) surface. Solid or broken curves are pseudo-Voigt peak fits. (d) A distorted $CuCl_6$ octahedron, observed quadrupole moments, $\mathbf{Q}_{\xi^2-\eta^2}$ (upper) and $\mathbf{Q}_{\eta\zeta}$ (lower), and a chiral quadrupole helix along [001] as a linear combination of the quadrupole moments. The helix is mirrored between two enantiomers (left: $P6_522$ and right: $P6_122$). Two colors of the quadrupole moments show the sign of poles, red (+) and blue (−). The local Cartesian coordinate system $\xi\eta\zeta$ is defined so that $\xi$ is along the two-fold axis (// <110>), $\zeta$ is along [001], and $\eta$ is normal to both of them. The quadrupole moment $\mathbf{Q}_{3\zeta^2-r^2}$, which is not observed in our experiment, is omitted here.

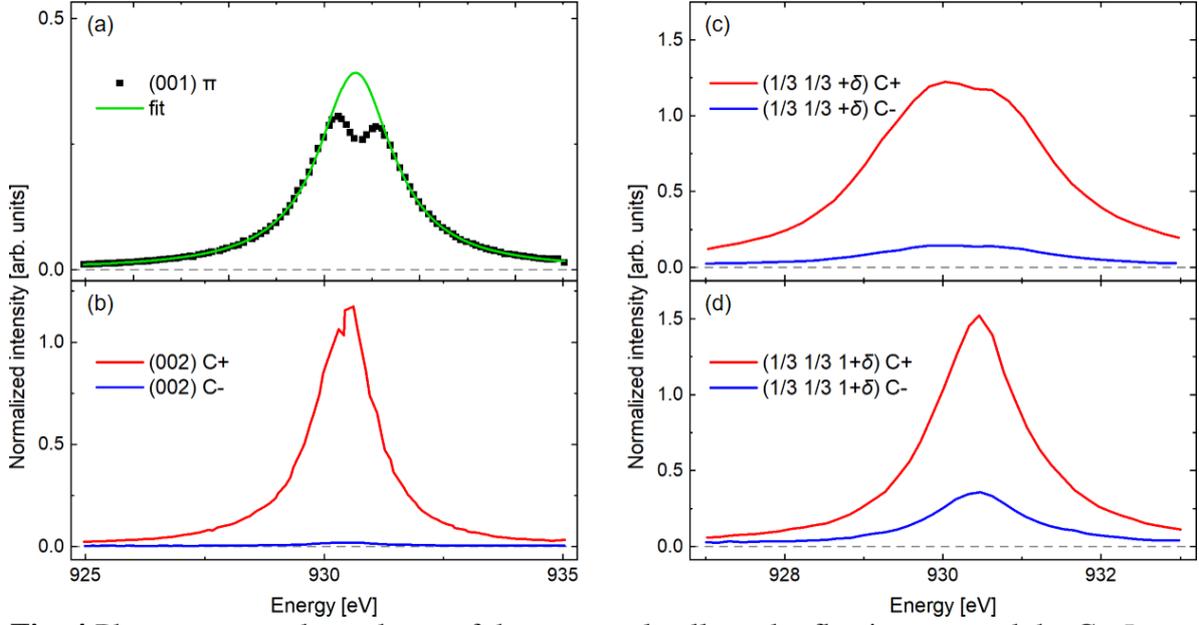

**Fig. 4** Photon-energy dependence of the resonantly allowed reflections around the Cu $L_3$ edge while maintaining a given diffraction condition; (a) (001) and (b) (002) forbidden reflections due to electric quadrupole moments from a left-handed ($P6_522$) crystal (sample#1) with the (001) surface, and (c) (1/3 1/3 +$\delta$) and (d) (1/3 1/3 1+$\delta$) magnetic reflections from sample#3 with the (110) surface. (c) and (d) were taken below $T_N$. A green curve in (a) is a Lorentzian fit to correct the self-absorption (see main text).

Figures 5(a) and 5(b) show RXD profiles of the (1/3 1/3 ±$\delta$) magnetic reflections from sample#3 with the (110) surface, whose resonant enhancement is shown in Fig. 4(c). The RXD intensities of magnetic reflections from the reported magnetic structure when using circularly polarized x-ray beams can be expressed as

$$I(h, P_2, \pm k) = \frac{I_G}{8}\{[\sin^2\omega + \sin^2 2\theta + \sin^2(2\theta - \omega)](\delta_{\tau,G+k} + \delta_{\tau,G-k}) + 2hP_2 \sin(2\theta - \omega)\sin 2\theta\,(\delta_{\tau,G+k} - \delta_{\tau,G-k})\}, \qquad (3)$$

where $\omega$ is the incident angle of x-ray beams to the (110) surface [see the inset of Fig. 5(a)], **G** represents a reciprocal lattice vector, **τ** is the scattering vector, and $h = -1$ (+1) indicates spin helicity for the left- (right-)handed magnetic structure {corresponding to Figs. 2(c) and 2(d) [Figs. 2(a) and 2(b)]} (See Appendix B for details). $I_G = F_G^* F_G$ is the diffraction intensity of a fundamental reflection at **G**. Here, $F_G = (3/4\pi q)(F_{-1}^1 - F_{+1}^1)\sum_j \exp(i\mathbf{G}\cdot\mathbf{r}_j)$ is the structure factor, where $q$ is the modulus of the wave vector of incident x-rays, $F_{\pm 1}^1$ is the atomic scattering properties of the electric-dipole transition, and $\mathbf{r}_j$ is the positional vector of the $j$th $Cu^{2+}$. Equation (3) relates the circular dichroism to the spin helicity. As the two propagation vectors $\mathbf{k}_1$ and $\mathbf{k}_2$ do not coexist in a single magnetic domain, the observation of two distinct magnetic reflections indicates the presence of configuration domains. A mono-chiral crystal exhibits a mono-chiral spin helix since the helical modulation results from the antisymmetric exchange interaction mediated through spin-orbit coupling [Kousaka2]. Thus, we obtain only two magnetic domains with a right-handed spin helix [Figs. 2(a) and 2(b)].

In addition to the observed (1/3 1/3 ±$\delta$) reflections, which are satellite reflections around **G** = (0, 0, 0), we observed the resonantly-allowed (1/3 1/3 1±$\delta$) reflections [see Figs. 4(d), 5(c), and 5(d)], exhibiting clear circular dichroism as well. These reflections are satellites around **G** = (0, 0, 1), which are absent for the reported magnetic structure because the (0 0 1) reflection is space-group forbidden, i.e., $I_{(001)} = 0$ in Eq. (3). This is consistent with the absence of intensity off-the resonance [Fig. 4(a)]. Nevertheless, their magnetic origin is evident because of the temperature dependence shown in Fig. 5(f) as they vanish above $T_N$.

To clarify their origin, we collected RXD data along (00$L$) and found a broad peak at (0 0 1/2), existing only below $T_N$ [see Fig. 5(e)]. The (0 0 1/2) reflection shows negligible circular dichroism and, therefore, probes another magnetic component than the helical component. The broad peak width indicates the small correlation length of this component that is ~4.5 nm, supporting that the component is independent of the helical one. Here the correlation length $p$ was obtained as $p = c/2\pi\Delta l$, where $\Delta l$ is the fitted half-width at half maximum of the reflection. Note that the penetration depth of the x-ray beams at the incidence angle for (0 0 1/2) estimated in the same way for (001) is ~52 nm. Considering (i) the resemblance to the reported magnetic signals in chiral langasite $Ba_3(Nb,Ta)Fe_3Si_2O_{14}$, (ii) the absence of circular dichroism, and (iii) no expected cycloidal component because of no electric polarization in the ground state [Miyake1], the additional component can be described by sinusoidal modulations along [001], as drawn in Fig. 1(d). Thus, this observation supports a longitudinal conical (or helical-butterfly) structure shown in Fig. 1(e). Indeed, the symmetry analysis using "k-Subgroupsmag" from the Bilbao Crystallographic Server [Perez-Mato1] gives such a magnetic structure as a possible magnetic subgroup of the space group of the paramagnetic phase with the given two magnetic propagation vectors $\mathbf{k}_{1(2)}$ and $\mathbf{k}_3 = (0, 0, 1/2)$.

In the presence of out-of-plane sinusoidal modulations with a propagation vector $\mathbf{k}_3$, the amplitude of the in-plane component modulates along [001] with a twice larger propagation vector than $\mathbf{k}_3$. This allows magnetic reflections at $\boldsymbol{\tau} = (0, 0, 0) + \mathbf{k}_{1(2)} + 2\mathbf{k}_3$, appearing at (1/3 1/3 1±$\delta$). Therefore, the RXD intensities of the family of (1/3 1/3 1±$\delta$) reflections can be written as

$$I(h, P_2) = \frac{I_G}{32} A_1^2 \{[\sin^2\omega + \sin^2 2\theta + \sin^2(2\theta - \omega)](\delta_{\boldsymbol{\tau},\mathbf{G}+\mathbf{k}_1+2\mathbf{k}_3} + \delta_{\boldsymbol{\tau},\mathbf{G}+\mathbf{k}_1-2\mathbf{k}_3} + \delta_{\boldsymbol{\tau},\mathbf{G}+\mathbf{k}_2+2\mathbf{k}_3} + \delta_{\boldsymbol{\tau},\mathbf{G}+\mathbf{k}_2-2\mathbf{k}_3}) + 2hP_2 \sin(2\theta - \omega)\sin 2\theta (-\delta_{\boldsymbol{\tau},\mathbf{G}+\mathbf{k}_1+2\mathbf{k}_3} - \delta_{\boldsymbol{\tau},\mathbf{G}+\mathbf{k}_1-2\mathbf{k}_3} + \delta_{\boldsymbol{\tau},\mathbf{G}+\mathbf{k}_2+2\mathbf{k}_3} + \delta_{\boldsymbol{\tau},\mathbf{G}+\mathbf{k}_2-2\mathbf{k}_3})\}, \quad (4)$$

where $A_1$ is a series expansion coefficient (see Appendix C for details). This matches well with the experimental observation, i.e., the emergence of the reflections with circular dichroism corresponding to spin helicity.

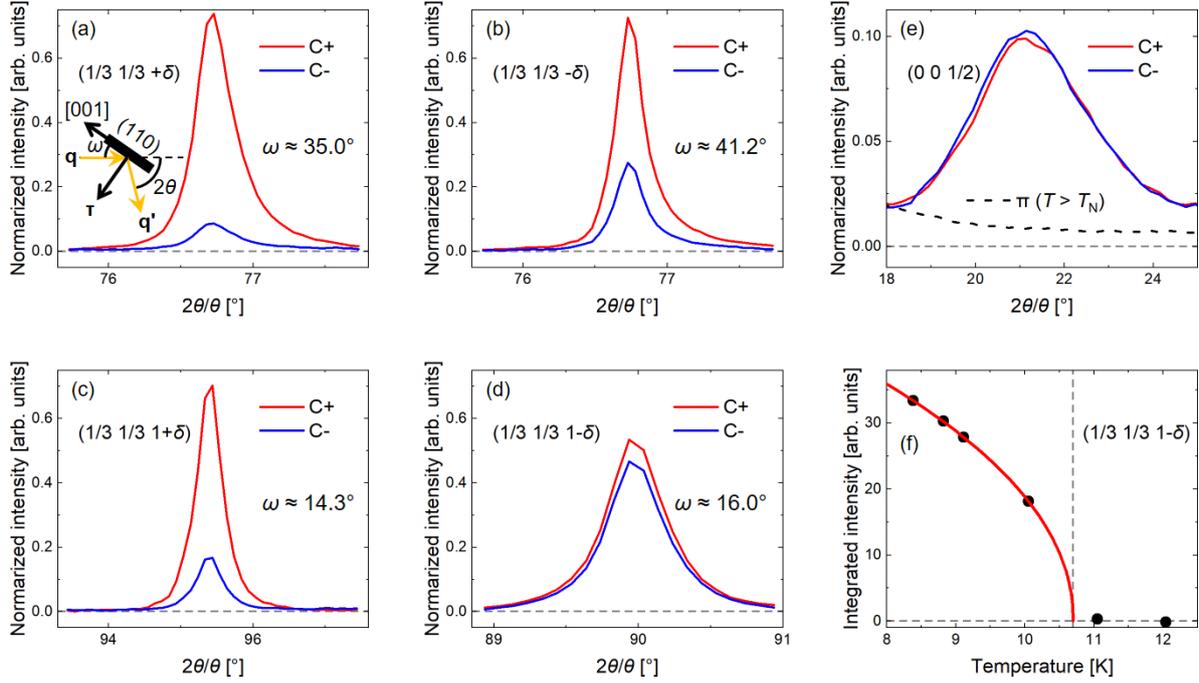

**Fig. 5** Resonant diffraction profiles of magnetic reflections measured below $T_N$; (a) (1/3 1/3 +$\delta$) [**G** = **0** + **k**$_1$], (b) (1/3 1/3 –$\delta$) [**G** = **0** + **k**$_2$], (c) (1/3 1/3 1+$\delta$) [**G** = **0** + 2**k**$_3$ + **k**$_1$], (d) (1/3 1/3 1–$\delta$) [**G** = **0** + 2**k**$_3$ + **k**$_2$], and (e) (0 0 1/2) [**G** = **0** + **k**$_3$]. For comparison, a profile measured above $T_N$ with the $\pi$ polarization is shown in (e) by a black broken line. The inset of (a) shows the diffraction geometry, where **q** (**q**') is the wave vector of incident (scattered) x-ray beam and **τ** is the scattering vector. (f) Temperature dependence of the (1/3 1/3 1–$\delta$) reflection from sample#2 with the (119) surface. Red curve in (f) represents power-law fit [$\propto (T_N - T)^\alpha$], where $\alpha$ (= 0.48 ± 0.03) is the critical exponent and $T_N$ is fixed to 10.7 K.

    Whereas the antisymmetric exchange interaction was proposed to create the sinusoidal modulation in a helically twisted 120° AFM structure for langasite Ba$_3$NbFe$_3$Si$_2$O$_{14}$ [Scagnoli1], this mechanism is unlikely applicable for CsCuCl$_3$ because the sinusoidal modulation has a different propagation vector than the helical component. The commensurate propagation vector implies its origin in local spin-orbit interaction, i.e., single-ion anisotropy. Although single-ion anisotropy has long been believed not to be relevant in $S = 1/2$ systems [Moriya2], its importance for such systems was pointed out by Liu *et al*. [Liu1]. Taking the quantization axis $z$ along the elongated direction of CuCl$_6$ octahedron, a hole populates $3d_{x^2-y^2}$ [Laiho1]. Our RXD results support this picture as the negative poles (electron) of the electric quadrupole moment of Cu $3d$ point to the $z$ axis while the positive poles (hole) point to the orthogonal directions, as shown in Fig. 3(d). The $z$ axis lies not in the basal plane, implying that single-ion anisotropy favors spins to point off the basal plane. This additional term in the magnetic Hamiltonian may stabilize the longitudinal conical structure.

    It might be worth comparing the longitudinal conical structure also with double molybdate RbFe(MoO$_4$)$_2$ because of similarity and difference, which exhibits a 120° AFM structure with helical modulation along [001] without a sinusoidal component [Marty1]. While there is orbital angular momentum $L$ in Fe$^{3+}$ for Ba$_3$NbFe$_3$Si$_2$O$_{14}$ due to strong hybridization between Fe $3d$ and O $2p$ orbitals [Scagnoli1] and in Cu$^{2+}$ for CsCuCl$_3$ as here

discussed, *L* may be negligible in $Fe^{3+}$ for $RbFe(MoO_4)_2$ as the bond length between Fe and O is much larger in $RbFe(MoO_4)_2$ than in $Ba_3NbFe_3Si_2O_{14}$ (more than 1 pm) [Jain1]. A negligibly small *L* results in a minor single-ion anisotropy insufficient to stabilize the longitudinal conical structure.

In conclusion, we performed resonant soft x-ray diffraction on a chiral crystal $CsCuCl_3$ and characterized its multi-chiral structures, i.e., the orbital chirality in the crystal structure and the magnetic structure. Two quadrupole moment components of the $Cu^{2+}$ 3*d* states, determined by the distorted $CuCl_6$ octahedron, were quantified by measuring two independent forbidden reflections. The result agrees with the presence of a hole in a specific 3*d* state and a chiral arrangement of the orbitals. In addition to the magnetic satellite reflections already observed by neutron diffraction originated from the 120° antiferromagnetic structure in the basal plane with a helical modulation along the principal axis, we found additional magnetic reflections implying the presence of sinusoidal modulations along the principal axis in the magnetic structure, i.e., a longitudinal conical (or helical-butterfly) structure. The out-of-plane sinusoidal modulations might be caused by a single-ion anisotropy with its local quantization axis of Cu 3*d* states being off the basal plane. Our results suggest a strong correlation between orbital and magnetism even in $S = 1/2$ systems and its importance to understanding the magnetic ground state.

**Acknowledgements**

We would like to thank Y. Kousaka for stimulating discussion and B. Roessli for helpful discussions on the symmetry analysis. The static resonant x-ray diffraction experiments were performed at the X11MA beamline in the Swiss Light Source under proposal Nos. 20200683 and 20210334 and in-house access. H.U. and V.U. acknowledges the National Centers of Competence in Research in Molecular Ultrafast Science and Technology (NCCR MUST-No. 51NF40-183615) and E.S. for the NCCR Materials' Revolution: Computational Design and Discovery of Novel Materials (NCCR MARVEL) from the Swiss National Science Foundation. H.U. and E.S. are also supported by the European Union's Horizon 2020 research and innovation programme under the Marie Skłodowska-Curie Grant Agreement No. 801459 – FP-RESOMUS and No. 884104 – PSI-FELLOW, respectively. M.B., G.P.S., and L.L. acknowledge funding from the Swiss National Science Foundation through project Nos. 200021-196964 and 200021_169698.

**Appendix A: Symmetry analysis and RXD intensities of forbidden reflections**

Here, we calculate RXD intensities of the forbidden reflections with referring to Refs. [Lovesey1,Nakajima1]. The electron density $\rho(\mathbf{r})$ around an ion can be expressed by electric multipoles $\rho_{lm}(\mathbf{r})$ as $\rho(\mathbf{r}) = \sum_{l,m} \rho_{lm}(r) Y_l^m(\hat{\mathbf{r}})$, where $\hat{\mathbf{r}}$ is a radial unit vector and $Y_l^m(\hat{\mathbf{r}})$ is the spherical harmonics with $-l \leq m \leq l$. We get $\rho_{lm}(r)$ as $\rho_{lm}(r) = \int \rho(\mathbf{r}) Y_l^m(\hat{\mathbf{r}})^* d\hat{\mathbf{r}}$. The multipole moments are generally expressed as an expected value of the spherical tensor $T_Q^K$, which relates to $Y_l^m(\hat{\mathbf{r}})$ as $T_Q^K = Y_l^m(\hat{\mathbf{r}})$ with $m = Q$ and $l = K$. Here, *K* is the rank of the tensor and *Q* is its projection, holding the relation $-K \leq Q \leq K$. $\langle T_Q^K \rangle$ is a complex number, $\langle T_Q^K \rangle = \langle T_Q^K \rangle' + i\langle T_Q^K \rangle''$, with $\langle T_Q^K \rangle^* = (-1)^Q \langle T_{-Q}^K \rangle$. There are five independent real-number components for quadrupole moments with $K = 2$, $\langle T_0^2 \rangle'$, $\langle T_{+1}^2 \rangle'$, $\langle T_{+1}^2 \rangle''$, $\langle T_{+2}^2 \rangle'$, and $\langle T_{+2}^2 \rangle''$, which corresponds to $Q_{3z^2-r^2}$, $Q_{zx}$, $Q_{yz}$, $Q_{x^2-y^2}$, and $Q_{xy}$ for a general Cartesian coordinate system $xyz$, respectively.

There are six $Cu^{2+}$ located at $\mathbf{r}_1 \sim \mathbf{r}_6$ in a single unit cell with the Wyckoff position $6a$, as listed in Table A1. Using $\langle T_Q^K \rangle$ of $Cu^{2+}$ at $\mathbf{r}_1$, those of the remaining five $Cu^{2+}$ can be obtained by rotating $\langle T_Q^K \rangle$ by $\chi\pi/3$ ($\mathbf{r}_2$), $2\chi\pi/3$ ($\mathbf{r}_3$), $\chi\pi/2$ ($\mathbf{r}_4$), $-2\chi\pi/3$ ($\mathbf{r}_5$), and $-\chi\pi/3$ ($\mathbf{r}_6$), where $\chi = -1$ (+1) corresponds to a left- (right-)handed crystal structure. The RXD structure factor $\Psi_Q^K$ of a (00L) reflection is

$$\Psi_Q^K = \langle T_Q^K \rangle \left(1 + e^{2\pi\chi i\frac{Q}{6}}e^{2\pi i\frac{L}{6}} + e^{2\pi\chi i\frac{Q}{3}}e^{2\pi i\frac{L}{3}} + e^{2\pi\chi i\frac{Q}{2}}e^{2\pi i\frac{L}{2}} + e^{2\pi\chi i\frac{Q}{3}}e^{-2\pi i\frac{L}{3}} + e^{2\pi\chi i\frac{Q}{6}}e^{-2\pi i\frac{L}{6}}\right).$$

(A1)

It is evident that a left- (right-)handed structure gives forbidden reflections when $L - Q = 6n$ ($L + Q = 6n$), where $n$ is an integer. Note that $\langle T_0^K \rangle$ does not contribute to the space-group forbidden reflections but to allowed reflections. We take the local Cartesian coordinate system $\xi\eta\zeta$ shown in Fig. 3(d), where $\xi$ is along the two-fold axis <110>, $\zeta$ is along [001], $\eta$ is normal to both the directions. The two-fold ($C_2$) symmetry constraints $\langle T_Q^K \rangle$ that is an odd function of $\xi$ to be zero, i.e., $\langle T_{+1}^2 \rangle'$ and $\langle T_{+2}^2 \rangle''$, corresponding to $\zeta\xi$ and $\xi\eta$, respectively. As a short summary, only one quadrupole moment contributes to respective forbidden (00L) reflections; $\langle T_{+1}^2 \rangle''$ ($\zeta\xi$) for (001) and $\langle T_{+2}^2 \rangle'$ ($\xi^2 - \eta^2$) for (002).

The scattering length at an atomic resonance is sensitive to the polarization of incident x-rays $\boldsymbol{\varepsilon}$ and that of scattered x-rays $\boldsymbol{\varepsilon}'$. The resonant scattering is then sensitive to anisotropic electron density characterized by electric quadrupole moments. An x-ray susceptibility tensor

$$\hat{f} = \begin{pmatrix} f_{\xi\xi} & f_{\xi\eta} & f_{\xi\zeta} \\ f_{\xi\eta} & f_{\eta\eta} & f_{\eta\zeta} \\ f_{\xi\zeta} & f_{\eta\zeta} & f_{\zeta\zeta} \end{pmatrix}, \qquad (A2)$$

defined by the local symmetry of a resonant atom, describes the scattering. We here take the local Cartesian coordinate system $\xi\eta\zeta$. Note that the tensor components and electric quadrupole moments are related as $Q_{3\zeta^2-r^2} = \frac{1}{2}(2f_{\zeta\zeta} - f_{\xi\xi} - f_{\eta\eta})$, $Q_{\xi\zeta} = \frac{2}{\sqrt{3}}f_{\xi\zeta}$, $Q_{\eta\zeta} = \frac{2}{\sqrt{3}}f_{\eta\zeta}$, $Q_{\xi^2-\eta^2} = \frac{1}{\sqrt{3}}(f_{\xi\xi} - f_{\eta\eta})$, and $Q_{\xi\eta} = \frac{2}{\sqrt{3}}f_{\xi\eta}$ [Nagao1]. The local $C_2$ symmetry along the $\xi$ axis requires the relation $\hat{f} = C_2 \hat{f} C_2^{-1}$, which results in $f_{\xi\eta} = f_{\xi\zeta} = 0$,

$$\hat{f} = \begin{pmatrix} f_{\xi\xi} & 0 & 0 \\ 0 & f_{\eta\eta} & f_{\eta\zeta} \\ 0 & f_{\eta\zeta} & f_{\zeta\zeta} \end{pmatrix}. \qquad (A3)$$

Each $Cu^{2+}$ position is connected by the six-fold screw symmetry along $\zeta$, whose $\hat{f}$ is thus obtained as shown in Table A1. The RXD form factor $\hat{F}$ from a single unit cell at the scattering vector $\boldsymbol{\tau} = (0, 0, L)$ is calculated as

$$\hat{F}_{(00L)} = \hat{f} + C_3^{\chi}\hat{f}C_3^{-\chi}e^{2\pi i\frac{L}{6}} + C_3^{2\chi}\hat{f}C_3^{-2\chi}e^{2\pi i\frac{L}{3}} + C_2^1\hat{f}C_2^{-1}e^{2\pi i\frac{L}{2}} + C_3^{-2\chi}\hat{f}C_3^{\chi}e^{-2\pi i\frac{L}{3}} +$$
$$C_3^{-\chi}\hat{f}C_3^{\chi}e^{-2\pi i\frac{L}{6}}. \quad (A4)$$

By using resonant scattering amplitude in the respective polarization channel $\varepsilon'\varepsilon$ ($\hat{F}_{(00L)}^{\varepsilon'\varepsilon}$) described as

$$\hat{F}_{(00L)}^{\varepsilon'\varepsilon} = \varepsilon'\hat{F}_{(00L)}\varepsilon, \quad (A5)$$

the RXD intensity $I_{(00L)}$ using circular polarization with the Stokes parameter $P_2$ is obtained as

$$I_{(00L)}(P_2) = \frac{1}{2}\left(\left|\hat{F}_{(00L)}^{\sigma'\sigma}\right|^2 + \left|\hat{F}_{(00L)}^{\pi'\sigma}\right|^2 + \left|\hat{F}_{(00L)}^{\sigma'\pi}\right|^2 + \left|\hat{F}_{(00L)}^{\pi'\pi}\right|^2\right) + P_2\text{Im}\left(\hat{F}_{(00L)}^{\sigma'\pi}{}^*\hat{F}_{(00L)}^{\sigma'\sigma} + \hat{F}_{(00L)}^{\pi'\pi}{}^*\hat{F}_{(00L)}^{\pi'\sigma}\right), \quad (A6)$$

while $I_{(00L)}$ using linear polarization with the Stokes parameter $P_3$ (+1 for $\sigma$ while –1 for $\pi$) is

$$I_{(00L)}(P_3) = \frac{1}{2}(1 + P_3)\left(\left|\hat{F}_{(00L)}^{\sigma'\sigma}\right|^2 + \left|\hat{F}_{(00L)}^{\pi'\sigma}\right|^2\right) + \frac{1}{2}(1 - P_3)\left(\left|\hat{F}_{(00L)}^{\sigma'\pi}\right|^2 + \left|\hat{F}_{(00L)}^{\pi'\pi}\right|^2\right). \quad (A7)$$

We obtain

$$I_{(001)}(P_3 = \pm 1) = I_{(001)}(P_2 = \pm 1) = 9f_{\eta\zeta}^2\cos^2\theta = \frac{27}{4}|Q_{\eta\zeta}|^2\cos^2\theta, \quad (A8)$$

$$I_{(002)}(\chi, P_2) = \frac{9}{8}(f_{\xi\xi} - f_{\eta\eta})^2(1 + \sin^2\theta)(1 - \chi P_2 \sin\theta)^2$$
$$= \frac{27}{8}|Q_{\xi^2-\eta^2}|^2(1 + \sin^2\theta)(1 - \chi P_2 \sin\theta)^2, \quad (A9)$$

$$I_{(002)}(P_3 = +1) = \frac{9}{8}(f_{\xi\xi} - f_{\eta\eta})^2(1 + \sin^2\theta)$$
$$= \frac{27}{8}|Q_{\xi^2-\eta^2}|^2(1 + \sin^2\theta), \text{ and} \quad (A10)$$

$$I_{(002)}(P_3 = +1) = \frac{9}{8}(f_{\xi\xi} - f_{\eta\eta})^2(1 + \sin^2\theta)\sin^2\theta$$
$$= \frac{27}{8}|Q_{\xi^2-\eta^2}|^2(1 + \sin^2\theta)\sin^2\theta, \quad (A11)$$

where $\theta$ is the Bragg angle. We find (002) shows circular dichroism correlating to crystal chirality while (001) does not. Unlike trigonal systems [Tanaka1,Usui1], there is no azimuthal angle dependence on the RXD intensities because such dependence appears due to a coupled term between two quadrupole moments.

Table A1 The atomic position, multipole moments, and x-ray susceptibility tensor of six $Cu^{2+}$ in a single unit cell. Here $\chi$ represents the crystal chirality, –1 (+1) for a left- (right-)handed structure.

| Label | Position | Multipole moments | X-ray susceptibility tensor |
|---|---|---|---|
| 1 | $\mathbf{r}_1 = (0, 0, 0)$ | $\langle T_Q^K \rangle$ | $\hat{f}$ |
| 2 | $\mathbf{r}_2 = \left(x, x, \frac{1}{6}\chi\right)$ | $\langle T_Q^K \rangle e^{2\pi\chi i\frac{Q}{6}}$ | $C_3^{\chi}\hat{f}C_3^{-\chi}$ |
| 3 | $\mathbf{r}_3 = \left(0, x, \frac{1}{3}\chi\right)$ | $\langle T_Q^K \rangle e^{2\pi\chi i \frac{Q}{3}}$ | $C_3^{2\chi} \hat{f} C_3^{-2\chi}$ |
| 4 | $\mathbf{r}_4 = \left(-x, 0, \frac{1}{2}\right)$ | $\langle T_Q^K \rangle e^{2\pi\chi i \frac{Q}{2}}$ | $C_2^{1} \hat{f} C_2^{-1}$ |
| 5 | $\mathbf{r}_5 = \left(-x, -x, \frac{1}{3}\chi\right)$ | $\langle T_Q^K \rangle e^{-2\pi\chi i \frac{Q}{3}}$ | $C_3^{-2\chi} \hat{f} C_3^{\chi}$ |
| 6 | $\mathbf{r}_6 = \left(0, -x, \frac{1}{6}\chi\right)$ | $\langle T_Q^K \rangle e^{-2\pi\chi i \frac{Q}{6}}$ | $C_3^{-\chi} \hat{f} C_3^{\chi}$ |

**Appendix B: RXD intensities of magnetic reflections from the helical structure**

The magnetic scattering term in the resonant scattering length from a single atom is

$$f_\mathrm{m} = -\left(\frac{3}{4\pi q}\right) i (\boldsymbol{\varepsilon}' \times \boldsymbol{\varepsilon}) \cdot \mathbf{m}(F_{-1}^1 - F_{+1}^1), \quad \text{(B1)}$$

where $\mathbf{m}$ is the unit vector along a magnetic moment, $q$ is the modulus of the wave vector of incident x-ray beams, and $F_{\pm 1}^1$ represents the atomic scattering properties of the dipole transition [Lovesey2]. We here use the Cartesian coordinate system *xyz*, where *x* is along [110], *y* is along [–210], and *z* is along [001] [see Fig. 5(a) for the diffraction geometry, an incident angle of $\omega$ and scattering angle of $2\theta$]. The photon polarization dependence ($\boldsymbol{\varepsilon}' \times \boldsymbol{\varepsilon}$) is given by

$$\boldsymbol{\varepsilon}' \times \boldsymbol{\varepsilon} = \begin{pmatrix} \boldsymbol{\sigma}' \times \boldsymbol{\sigma} & \boldsymbol{\sigma}' \times \boldsymbol{\pi} \\ \boldsymbol{\pi}' \times \boldsymbol{\sigma} & \boldsymbol{\pi}' \times \boldsymbol{\pi} \end{pmatrix} = \begin{pmatrix} 0 & \hat{\mathbf{q}} \\ -\hat{\mathbf{q}}' & \hat{\mathbf{q}}' \times \hat{\mathbf{q}} \end{pmatrix}, \quad \text{(B2)}$$

where $\hat{\mathbf{q}} = (-\sin\omega, 0, -\cos\omega)$ [$\hat{\mathbf{q}}' = (\sin(2\theta - \omega), 0, -\cos(2\theta - \omega))$] is the unit vector along the wave vector of incident [scattered] x-ray beams and $\hat{\mathbf{q}}' \times \hat{\mathbf{q}} = (0, \sin 2\theta, 0)$. Total scattering amplitude $F$ is described by using a magnetic form factor $\mathbf{F}_m = \sum_j \mathbf{m}_j e^{i\boldsymbol{\tau}\cdot\mathbf{r}_j}$ and $b = -\left(\frac{3}{4\pi q}\right) i (F_{-1}^1 - F_{+1}^1)$ as

$$F = b \begin{pmatrix} 0 & \hat{\mathbf{q}} \cdot \mathbf{F}_m \\ -\hat{\mathbf{q}}' \cdot \mathbf{F}_m & (\hat{\mathbf{q}}' \times \hat{\mathbf{q}}) \cdot \mathbf{F}_m \end{pmatrix}. \quad \text{(B3)}$$

The *j*th $Cu^{2+}$ in the helical magnetic structure of $CsCuCl_3$ has the magnetic moment

$$\mathbf{m}_j = \begin{pmatrix} \cos(i\mathbf{k}\cdot\mathbf{r}_j) \\ \sin(i\mathbf{k}\cdot\mathbf{r}_j) \\ 0 \end{pmatrix} = \frac{1}{2}\begin{pmatrix} e^{i\mathbf{k}\cdot\mathbf{r}_j} + e^{-i\mathbf{k}\cdot\mathbf{r}_j} \\ -ih(e^{i\mathbf{k}\cdot\mathbf{r}_j} - e^{-i\mathbf{k}\cdot\mathbf{r}_j}) \\ 0 \end{pmatrix}. \quad \text{(B4)}$$

Here $\mathbf{r}_j$ is the positional vector of the *j*th $Cu^{2+}$, $\mathbf{k}$ is the magnetic propagation vector, either $\mathbf{k}_1$ or $\mathbf{k}_2$, and $h = -1$ (+1) describes the spin helicity of a left- (right-)handed helical magnetic structure. $\mathbf{F}_m$ is calculated by summing up $\mathbf{m}_j$ at all positions in a crystal with a phase factor,

$$\mathbf{F}_m = \sum_j \mathbf{m}_j e^{i\boldsymbol{\tau}\cdot\mathbf{r}_j} = \frac{F_G}{2}\begin{pmatrix} \delta_{\boldsymbol{\tau},\mathbf{G}-\mathbf{k}} + \delta_{\boldsymbol{\tau},\mathbf{G}+\mathbf{k}} \\ ih(\delta_{\boldsymbol{\tau},\mathbf{G}-\mathbf{k}} - \delta_{\boldsymbol{\tau},\mathbf{G}+\mathbf{k}}) \\ 0 \end{pmatrix}, \quad \text{(B5)}$$

where $\mathbf{G}$ is a reciprocal lattice vector and $F_{\mathbf{G}} = \sum_j e^{i\mathbf{G}\cdot\mathbf{r}_j}$ is the crystal structure factor for the scattering vector $\mathbf{\tau} = \mathbf{G}$. Using Eqs. (A6), (B3), and (B5), RXD intensity of magnetic reflections from the helical magnetic structure when using circular polarization is obtained as

$$I(h, P_2) = \frac{I_G}{8}\{[\sin^2\omega + \sin^2 2\theta + \sin^2(2\theta - \omega)](\delta_{\tau,\mathbf{G}+\mathbf{k}} + \delta_{\tau,\mathbf{G}-\mathbf{k}}) + 2hP_2 \sin(2\theta - \omega) \sin 2\theta \,(\delta_{\tau,\mathbf{G}+\mathbf{k}} - \delta_{\tau,\mathbf{G}-\mathbf{k}})\}. \quad (B6)$$

Here, $I_G = F_G^* F_G$ gives the intensity of the fundamental reflection at the scattering vector $\mathbf{\tau} = \mathbf{G}$. Equation (B6) explains the magnetic satellite reflections around $\mathbf{G} = (0, 0, 0)$ with circular dichroism correlating to $h$, i.e., $(1/3\ 1/3\ \pm\delta)$, while does not for those around $\mathbf{G} = (0, 0, 1)$, i.e., $(1/3\ 1/3\ 1\pm\delta)$, as $(001)$ is a forbidden reflection.

**Appendix C: RXD intensities of magnetic reflections with sinusoidal modulations**

With the presence of the sinusoidal modulations along [001] described by $\mathbf{k}_3$, the in-plane amplitude of the helical component modulates along [001] with a twice larger wave vector than $\mathbf{k}_3$. The in-plane amplitude for the $j$th $Cu^{2+}$ can be expanded as $A_0 + A_1 \cos(2\mathbf{k}_3 \cdot \mathbf{r}_j) + \cdots$, where $A_i$ is the $i$th coefficient of series expansion. Hence, $\mathbf{m}_j$ is written as

$$\mathbf{m}_j = A_0 \begin{pmatrix} \cos(i\mathbf{k}\cdot\mathbf{r}_j) \\ \sin(i\mathbf{k}\cdot\mathbf{r}_j) \\ 0 \end{pmatrix} + A_1 \begin{pmatrix} \cos(2\mathbf{k}_3\cdot\mathbf{r}_j)\cos(i\mathbf{k}\cdot\mathbf{r}_j) \\ \cos(2\mathbf{k}_3\cdot\mathbf{r}_j)\sin(i\mathbf{k}\cdot\mathbf{r}_j) \\ 0 \end{pmatrix} + \cdots + \begin{pmatrix} 0 \\ 0 \\ \Delta\sin(\mathbf{k}_3\cdot\mathbf{r}_j) \end{pmatrix}$$

$$= \frac{A_0}{2} \begin{pmatrix} e^{i\mathbf{k}\cdot\mathbf{r}_j} + e^{-i\mathbf{k}\cdot\mathbf{r}_j} \\ -ih(e^{i\mathbf{k}\cdot\mathbf{r}_j} - e^{-i\mathbf{k}\cdot\mathbf{r}_j}) \\ 0 \end{pmatrix} +$$

$$\frac{A_1}{4} \begin{pmatrix} e^{i(2\mathbf{k}_3+\mathbf{k})\cdot\mathbf{r}_j} + e^{-i(2\mathbf{k}_3+\mathbf{k})\cdot\mathbf{r}_j} + e^{i(2\mathbf{k}_3-\mathbf{k})\cdot\mathbf{r}_j} + e^{-i(2\mathbf{k}_3-\mathbf{k})\cdot\mathbf{r}_j} \\ -ih[e^{i(2\mathbf{k}_3+\mathbf{k})\cdot\mathbf{r}_j} - e^{-i(2\mathbf{k}_3+\mathbf{k})\cdot\mathbf{r}_j} - e^{i(2\mathbf{k}_3-\mathbf{k})\cdot\mathbf{r}_j} + e^{-i(2\mathbf{k}_3-\mathbf{k})\cdot\mathbf{r}_j}] \\ 0 \end{pmatrix} + \cdots -$$

$$i\frac{\Delta}{2} \begin{pmatrix} 0 \\ 0 \\ e^{i\mathbf{k}_3\cdot\mathbf{r}_j} - e^{-i\mathbf{k}_3\cdot\mathbf{r}_j} \end{pmatrix}, \quad (C1)$$

where $\Delta$ is the relative amplitude of the sinusoidal component with respect to the helical component without the modulations. Note that the coefficients, $A_i$ and $\Delta$, keep $|\mathbf{m}_j| = 1$. $\mathbf{F}_m$ is calculated as

$$\mathbf{F}_m = \frac{A_0}{2} F_G \begin{pmatrix} \delta_{\tau,\mathbf{G}-\mathbf{k}} + \delta_{\tau,\mathbf{G}+\mathbf{k}} \\ -ih(\delta_{\tau,\mathbf{G}-\mathbf{k}} - \delta_{\tau,\mathbf{G}+\mathbf{k}}) \\ 0 \end{pmatrix} +$$

$$\frac{A_1}{4} F_G \begin{pmatrix} \delta_{\tau,\mathbf{G}-2\mathbf{k}_3-\mathbf{k}} + \delta_{\tau,\mathbf{G}+2\mathbf{k}_3+\mathbf{k}} + \delta_{\tau,\mathbf{G}-2\mathbf{k}_3+\mathbf{k}} + \delta_{\tau,\mathbf{G}+2\mathbf{k}_3-\mathbf{k}} \\ -ih[\delta_{\tau,\mathbf{G}-2\mathbf{k}_3-\mathbf{k}} - \delta_{\tau,\mathbf{G}+2\mathbf{k}_3+\mathbf{k}} - \delta_{\tau,\mathbf{G}-2\mathbf{k}_3+\mathbf{k}} + \delta_{\tau,\mathbf{G}+2\mathbf{k}_3-\mathbf{k}}] \\ 0 \end{pmatrix} + \cdots -$$

$$i\frac{\Delta}{2} F_G \begin{pmatrix} 0 \\ 0 \\ \delta_{\tau,\mathbf{G}-\mathbf{k}_3} - \delta_{\tau,\mathbf{G}+\mathbf{k}_3} \end{pmatrix}. \quad (C2)$$

RXD intensities for the magnetic satellite reflections observed in our experiment are obtained as

$$I(h, P_2) = \frac{I_G}{8}\left\{[\sin^2\omega + \sin^2 2\theta + \sin^2(2\theta - \omega)]\left[A_0^2(\delta_{\tau,G-k} + \delta_{\tau,G+k}) + \frac{A_1^2}{4}(\delta_{\tau,G-2k_3-k} + \delta_{\tau,G+2k_3+k} + \delta_{\tau,G-2k_3+k} + \delta_{\tau,G+2k_3-k})\right] + 2\Delta^2[\cos^2(2\theta - \omega) + \cos^2\omega](\delta_{\tau,G-k_3} + \delta_{\tau,G+k_3}) + 2hP_2 \sin(2\theta - \omega)\sin 2\theta \left[A_0^2(\delta_{\tau,G-k} - \delta_{\tau,G+k}) + \frac{A_1^2}{4}(\delta_{\tau,G-2k_3-k} - \delta_{\tau,G+2k_3+k} - \delta_{\tau,G-2k_3+k} + \delta_{\tau,G+2k_3-k})\right]\right\}. \quad (C3)$$

Equation (C3) explains the appearance of magnetic satellite reflections around (001), i.e., (1/3 1/3 1±$\delta$) [$\tau$ = (0, 0, 1) + 2$k_3$ + $k$], and those due to the sinusoidal modulations (0 0 ±1/2) [$\tau$ = (0, 0, 0) + $k_3$], in addition to those around (000), i.e., (1/3 1/3 ±$\delta$). (1/3 1/3 1±$\delta$) show circular dichroism correlating to $h$ as similar to (1/3 1/3 ±$\delta$), whereas (0 0 ±1/2) does not.